\documentclass[twocolumn,showpacs,preprintnumbers,amsmath,amssymb]{revtex4}

\usepackage{graphicx}
\usepackage{dcolumn}
\usepackage{bm}

\begin{document}

\title{Optimal Monte Carlo Updating}

\author{Lode Pollet}
\email{Lode.Pollet@UGent.be}
\author{Stefan M.A. Rombouts}
\author{Kris Van Houcke}
\author{Kris Heyde}
\affiliation{Vakgroep subatomaire en stralingsfysica, \\
  Proeftuinstraat 86, \\
   Universiteit Gent, Belgium}

\date{\today}

\begin{abstract}
Based on Peskun's theorem it is shown that optimal transition matrices in
Markov chain Monte Carlo should have zero diagonal elements except for the
diagonal element corresponding to the largest weight. We will compare the
statistical efficiency of this sampler to existing algorithms, such as 
heat-bath updating and the Metropolis algorithm. We provide numerical results for
the Potts model as an application in classical physics. As an application in
quantum physics we consider the spin $3/2 XY$ model  and the Bose-Hubbard model
which have been simulated by the directed loop algorithm in the stochastic
series expansion framework.   
\end{abstract}

\maketitle
Monte Carlo methods are nowadays used in almost every branch of science,
offering exact results in a statistical sense or providing answers where other
methods fail. Already in statistical physics alone, Monte Carlo methods have
been applied to a variety of models~\cite{Newman99}. For many applications, good algorithms
have been devised and there exist now solutions to many problems that were
initially untractable. A well known example is  the critical slowing down in the
neighborhood of critical points that has been overcome in both classical (by
cluster algorithms~\cite{Swendsen87}~\cite{Wolff89}) and quantum Monte Carlo
(by the loop algorithms~\cite{Evertz93}). The need for better performing
algorithms is clear: efficient algorithms lead to more accurate results at the
same computational cost. Yet little that goes beyond common sense reasoning is
known about why an algorithm is efficient or not, and within a chosen
algorithm there is often additional freedom. 

We address the question of the efficiency of Markov chain Monte Carlo (MCMC) algorithms in terms
of smaller error bars. We first touch upon the needed terminology as it is usually~\cite{Evertz03}
understood in statistical physics from a practitioner's viewpoint. We show how
optimal sampling enters into this discussion and comment on its
implementation. Finally, we compare it with standard updating mechanisms for
the Potts model and for the directed loop algorithm~\cite{Syljuasen02, Syljuasen03} in the
stochastic series expansion~\cite{Sandvik99}. 

In MCMC a transition kernel (matrix) $T$ is set up and we will assume that we know the
discrete weights $W_1, \ldots , W_n$ (finite, computable set) of the invariant probability
distribution $W$. If the following two conditions hold\\
\begin{enumerate}
\item[(i)] normalization of probability, $\sum_j T_{ij} = 1, \forall i$
\item[(ii)] reversibility (detailed balance), $W_i T_{ij} = W_j T_{ji}$,
\end{enumerate}
and the chain can connect any two states in a certain finite number of steps,
then the Markov chain will converge to the invariant probability 
distribution (which will be $W$). The stochastic matrix $T$ has as largest eigenvalue $1$,
while the other eigenvalues are sorted by $-1 < \lambda_j < 1, j = 2, \ldots
n$. Strictly speaking, the condition (ii) is a too
strong~\cite{Tierney94}~\cite{Manousiouthakis99} condition to
assure convergence of the Markov chain towards the invariant distribution $W$,
it suffices $\sum_i W_i T_{ij} \sim W_j$, but the reversibility condition is
widely used in practical applications. 

The Markov process correlates the measurements of the observables $Q$ in
consecutive steps. The variance $\sigma^2_Q$ on these correlated
measurements is not equal to the variance $\sigma_{0,Q}^2$ obtained from
uncorrelated measurements. Instead, $\sigma^2_Q = 2\tau_{int, Q} \sigma_{0,Q}^2$, in
which we have introduced the integrated autocorrelation
time~\cite{Evertz03}~\cite{Janke01}, 
\begin{equation}\label{eq:tauint}
\tau_{int, Q} = \frac{1}{2}+ \sum_{t=1}^{\infty} A_Q(x^{(t)}).
\end{equation}
Stationary samples $x^{(t)}$ at the Monte Carlo times $t$ are obtained from
the sampler while the normalized autocorrelation function $A_Q(x^{(t)})$
for the observable $Q$ is given by 
\begin{equation}
A_Q(x^{(t)}) = \frac{\langle Q(x^{(i+t)}) Q(x^{(i)}) \rangle - \langle Q(x^{(i)})
  \rangle^2}{\langle Q^2(x^{(i)}) \rangle - \langle Q(x^{(i)}) \rangle^2}, 
\end{equation}
in which the ensemble average $\langle \ldots \rangle$ is taken over $i$. We
can now make a connection with the second largest eigenvalue by
\begin{equation}
\sup_Q \tau_{int_Q} = \frac{1+ \lambda_2}{2(1- \lambda_2)}.
\end{equation}
The following discussion focuses on the eigenvalues $\lambda_2, \lambda_3,
\ldots$ to obtain a lower asymptotic variance for an observable $Q$,
\begin{equation}
v(Q,T) = \lim_{n \to \infty} \frac{1}{n} \mbox{var} \left[ \sum_{k=1}^n Q(x^{(k)})
\right].\label{eq:asymptotic_var}
\end{equation}
A different question concerns the convergence~\cite{Mira98} of a probability distribution towards the invariant
probability distribution. It is dominated by the second largest
eigenvalue in absolute value of the stochastic matrix, which can be different
from $\lambda_2$ for non-positive operators, and would determine the required
number of thermalization or burn-in steps. Note that non-reversible transition
kernels can converge faster~\cite{Diaconis97}. 

The stochastic matrix has the dimension of the Hilbert space, and all
algorithms consist of two different operations in every Monte Carlo step: a limitation on
the configurations that can be reached and secondly the acceptance or the
rejection of the transition to one of them. For instance, heat-bath updating
(also called the Gibbs sampler~\cite{Geman84}) in the Ising model with
dimension $L \times L$ can in one step reach only $L^2+1$ 
different configurations from the current configuration which has weight
$W_1$. Among the $L^2$ new ones, it picks one at random and the transition to
this trial configuration with weight $W_2$ will be accepted according to
$\frac{W_2}{W_1 + W_2}$, otherwise it remains
in the current configuration. Note that all eigenvalues of the Gibbs sampler
are positive~\cite{Liu95}. 

We will now focus on this second step of the update. For the Ising model,
there are only two different configurations that play a role. How should we
choose the transition matrix so that the asymptotic variance is smallest? A
hint is given by the Peskun theorem~\cite{Peskun73}, stating that if $T^A$ and
$T^B$ both satisfy the conditions (i) and (ii) and if all off-diagonal
elements of $T^B$ are larger or equal than the corresponding elements of
$T^A$, then $T^B$ will lead to a smaller asymptotic variance for all
observables than $T^A$, or equivalently, $\lambda_2^A > \lambda_2^B$. It then
follows that the Metropolis-Hastings algorithm~\cite{Hastings70} is by
construction the most effective sampler for the Ising model with random single
site updates (and not the Gibbs sampler). For all possible stochastic matrices
with dimension $n=2$ the Metropolis transition matrix~\cite{Metropolis53} is
given by  
\begin{equation}
T_{ij}^{Met} = \left[
\begin{array}{cc}
0 & 1\\
\frac{W_1}{W_2} & 1-\frac{W_1}{W_2}
\end{array}
\right].\label{eq:met}
\end{equation}
Here we have ordered the weights in ascending order. This non-standard way of
writing Metropolis updating shows however the key ingredients of its efficiency,
namely that the chance of staying in the current configuration should be
minimized and secondly that the second largest eigenvalue is $\lambda_2 = -
T_{21}^{Met} = - \frac{W_1}{W_2}$.

Peskun's theorem implies an ordering of the weights. So let (for the remaining of
the paper) $\pi_1 \le \pi_2 \le \ldots \le \pi_n$ be the normalized weights in
ascending order, $\pi_i = \frac{W_i}{\sum_j W_j}$. Peskun's theorem tells us
that we can always improve a transition matrix $T$ by 'Metropolizing' it,
$T'_{ij} = \frac{T_{ij}}{\sum_{j\neq i} T_{ij}}, \forall j \neq i$. Applying
this idea to heat-bath updates, the following Metropolized Gibbs sampler(MG) is
obtained,  
\begin{equation}
T_{ij}^{MG} =  \left[ 
\begin{array}{ccccc}
0 & \frac{\pi_2}{1-\pi_1} & \frac{\pi_3}{1-\pi_1} & \ldots & \frac{\pi_n}{1-\pi_1}\\ 
\frac{\pi_1}{1-\pi_1} & 1-\ldots & \frac{\pi_3}{1-\pi_2} & \ldots & \frac{\pi_n}{1-\pi_2} \\
\frac{\pi_1}{1-\pi_1} & \frac{\pi_2}{1-\pi_2} & 1-\ldots & \ldots & \frac{\pi_n}{1-\pi_3}\\
\vdots & \vdots & \vdots & \ddots & \vdots \\
\frac{\pi_1}{1-\pi_1} & \frac{\pi_2}{1-\pi_2} & \frac{\pi_3}{1-\pi_3} & \ldots
& 1 - \ldots
\end{array}
\right],\label{eq:Liu}
\end{equation}
or $T_{ij}^{MG} = \min(\frac{\pi_j}{1-\pi_i},\frac{\pi_j}{1-\pi_j})$.
Liu~\cite{Liu96} has applied this idea to the independence Gibbs sampler, and
obtained a complete eigenanalysis for the resulting stochastic matrix.

It is possible to repeat this Metropolizing procedure until all but one of the
diagonal elements are zero. So, optimal transition matrices must have $T_{ii}
= 0, i \neq n$. Indeed, Frigessi {\it et al.}~\cite{Frigessi92} have shown
that the optimal transition matrix is of the form,
\begin{equation}
T_{ij}^{Opt} =  \left[ 
\begin{array}{ccccc}
0 & \frac{W_2}{W_1}y_1 & \frac{W_3}{W_1}y_1 & \ldots & \frac{W_n}{W_1}y_1\\ 
y_1 & 0 & \frac{W_3}{W_2}y_2 & \ldots & \frac{W_n}{W_2}y_2 \\
y_1 & y_2 & 0 & \ldots & \frac{W_n}{W_3}y_3\\
\vdots & \vdots & \vdots & \ddots & \vdots \\
y_1 & y_2 & y_3 & \ldots & 1 - y_1 - y_2 - \ldots
\end{array}
\right],\label{eq:stoch}
\end{equation}
with $y_1 = \frac{\pi_1}{1-\pi_1}, y_2 = (1-y_1)\frac{\pi_2}{1-\pi_1-\pi_2},
\ldots$. The eigenvalues are given by $1, \lambda_2 = -y_1 $ (the same as in Eq.(\ref{eq:Liu})) $, \lambda_3 = -y_2,
\ldots$. They are all negative and appear in an ordered way. This $\lambda_2$ has the lowest value
that possibly can be obtained with respect to the probability distribution
$W$, and with $\lambda_2$ determined, $\lambda_3$ is then the smallest
possible third largest eigenvalue, etc. Note that a rescaling is at work here, the entries for the second row 
$T_{2j}, j = 2, \ldots, n$ are analogous to the first row apart from the
rescaling $ 1 \to (1 - y_1)$.
Eq.(~\ref{eq:stoch}) represents an optimal transition matrix over the
entire Hilbert space, however, many situations of practical interest need to
sample stochastic subprocesses. Within these, optimal sampling can only be
achieved when all but one of the diagonal elements are zero.
When Eq.(~\ref{eq:stoch}) is applied to a stochastic subprocess, we will call it
the locally optimal algorithm.\\

{\it Potts model.} As a first application, we consider the $q=4$ Potts model~\cite{Newman99}
in two dimensions. We are interested in the dynamics of the Monte Carlo
process. Therefore we consider a small lattice with single-spin updates only,
and we do not want to use cluster updates~\cite{Swendsen87}~\cite{Wolff89} here. So we
will randomly select a spin, after which we have to make a choice between the
four possible orientations that this spin can take. Although the random selection of a site and
the single spin update both seriously violate the structure of the optimal
stochastic matrix Eq.(~\ref{eq:stoch}), the Peskun theorem still holds,
meaning that the matrix Eq.(~\ref{eq:stoch}) (applied to the stochastic
subprocess) still leads to a more effective sampling than heat-bath updating,
$T_{ij} = \frac{W_j}{\sum_k W_k} = \pi_j, \forall i,j$. This of course cannot
cure the fact that the updating of a single spin still leads to the divergence
of the autocorrelation time when the temperature reaches its critical value
$\beta_c$.

\begin{figure}
\includegraphics[width=8.5cm]{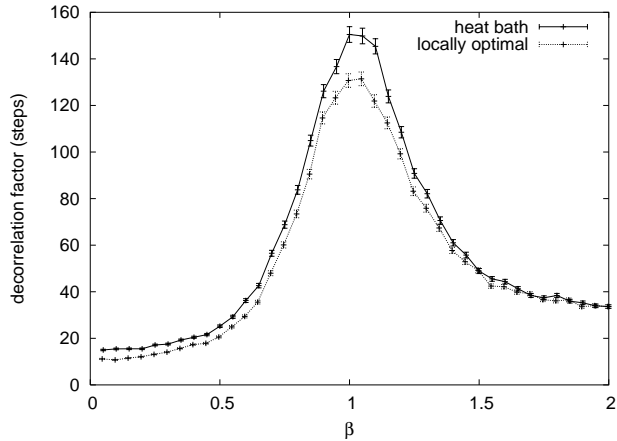}
\caption{\label{fig:potts} The decorrelation factor for the energy as a
  function of temperature $\beta$ at constant interaction strength for the
  $q=4$ Potts model on a $L = 4 \times 4$ matrix is shown for heat-bath
  updates and for the locally optimal ones. Simulations consisted of $4000$ chains of
  one million steps each.} 
\end{figure}

The decorrelation factor~\cite{Rombouts97} $\sigma_Q^2/\sigma_{0,Q}^2$ is
defined as the ratio between the error bars obtained from a Monte-Carlo
simulation with correlations between successive samples and the error bars one
would obtain from the same number of independent but identically distributed
samples. In the limit of a large number of samples, the decorrelation factor
becomes equal to twice the integrated autocorrelation time
(Eq.(\ref{eq:tauint})). The decorrelation factor can accurately be estimated
by running a large number of independent Markov chains. In
Fig.~\ref{fig:potts} we see that the decorrelation factor is much smaller for
the locally optimal algorithm than for the heat-bath algorithm.

A small lattice of $L = 4 \times 4$ has been chosen since the decorrelation
 factor scales with system size for single-site updates such that the    
 difference between the two algorithms will be larger in absolute terms when
 using a small lattice. We clearly see a significant difference between the
 locally optimal and the heat-bath transition probabilities. The Metropolized
 Gibbs sampler $T^{MG}$ (Eq.(~\ref{eq:Liu})) has the same $\lambda_2$ as the
 optimal one (Eq.(~\ref{eq:stoch})), and the integrated autocorrelation times
 differ only slightly for the $4 \times 4$ Potts model. Note that it is much
 easier to implement $T^{MG}$ than $T^{Opt}$, and if in practice the
 stochastic matrices cannot be computed in advance and need to be recomputed
 at each step, it is recommended to use $T^{MG}$. \\ 

{\it Directed Loops.} The same reasoning also applies to quantum Monte
Carlo. In the stochastic series expansion method~\cite{Sandvik99} a Taylor expansion
is applied to the partition function $Z=Tr \exp(-\beta H)$, yielding
\begin{eqnarray}
Z & = &  \sum_{m=0}^{\infty} \frac{\beta^m}{m!} \sum_{\{i_1,\ldots, i_m\}}
\sum_{\{b_1,\ldots,b_m\}} \langle i_1\vert -H_{b_1} \vert i_2 \rangle \langle
i_2 \vert -H_{b_2} \vert i_3 \rangle \nonumber \\ 
{} & {} & \cdots \langle i_{m-1} \vert -H_{b_{m-1}}\vert
i_m \rangle \langle i_m \vert -H_{b_m}\vert
i_1 \rangle,\label{eq:exp}
\end{eqnarray}
where the Hamiltonian $H$ is decomposed in a set of bond operators, $H =
\sum_b H_b$, and complete sets $ \vert i_k \rangle, k = 1, \ldots, n$ have
been inserted. In the first step, a diagonal update is performed in which the
expansion order $m$ can change. The second, off-diagonal update mimics the idea
of loop-type~\cite{Evertz93} and worm~\cite{Prokofev96} updates and can best be explained using a
graphical interpretation. Every matrix element in Eq.(~\ref{eq:exp}) is called
a vertex, at which the two sites of the interaction and an imaginary time are
assigned. Every vertex has four legs: two incoming and two outgoing legs per
site,  corresponding to the particles created and annihilated by $H_b$. The
legs of the vertices are connected by segments corresponding to the occupied sites. A worm is 
created in a arbitrary point in space-time by inserting a creation and
annihilation operator on a segment. One of these operators is chosen to be the worm head
and is mobile, while the other one is the worm tail and remains immobile. The
worm head moves through configuration space and can change the type of the
vertices, for instance from a diagonal to an off-diagonal vertex. The worm
movement stops when the worm head bites into its own tail. The entire movement
of the worm can then be regarded as a single loop update.

\begin{figure}
\includegraphics[width=8.5cm]{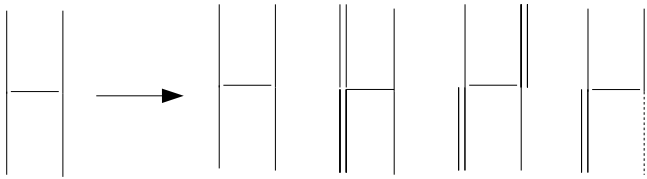}
\caption{\label{fig:scat} The four possible states that can arise when the
  worm enters the leftmost vertex at the leg left under in a number occupation
  basis (can also be a spin state) and for a system with particle number conservation. A single line
  means that the leg is singly occupied, a double line means double occupancy
  and a dashed line denotes that the leg is not occupied. The four processes
  on the right correspond to bounce, straight, jump and turn, from left to right.}
\end{figure}

Once the worm has been created, its movement is completely determined by how
it passes through and modifies the vertices. Suppose the worm head reaches a
vertex at the leg left under (entrance leg), as in Fig.~\ref{fig:scat}. The
local configuration can then change according to the four processes bounce,
straight, jump and turn, each with their proper weight $W_i, i =
1,\ldots,n$. For models with number conservation, it is possible that one or
more of the four processes cannot occur, so $n$ can in principle be two, three
or four. The bounce process is always possible but since it does not change
the current vertex, it can be regarded as a waste of computer time. The worm head has to
choose between one of the $n$ processes, modifies hereby the current vertex and
goes consequently to the next vertex along the segment that connects the
current exit leg and the next entrance leg. The probability matrix $T_{ij}$
defines the transition from the entrance leg to the exit legs and hence
completely determines the worm movement. We will now discuss several choices
for this probability matrix $T_{ij}$.

Originally, the heat-bath updates~\cite{Sandvik99}, $T_{ij} =
\frac{W_i}{\sum_j W_j}$, (solution A) were proposed.
Secondly, other choices are perfectly possible:
Sylju{\aa}sen and Sandvik propose {\em directed
loops}~\cite{Syljuasen02}, where the worm head has a preferred
direction at the vertices in order to be more
efficient than solution A. The rule of
thumb is that the frequency of the bounce processes should be as low as possible. This inspired
the authors of Ref.~\cite{Alet03} to numerically minimize the trace of the probability
matrix $T_{ij}$ with respect to detailed balance, $W_i T_{ij} = W_j T_{ji}$
(hereafter called solution B or the minimal bounce solution). They used a
linear programming technique~\cite{NumRec} for this purpose. Equivalently, one
can say that this amounts to minimizing the sum of the eigenvalues of the
transition matrix $T_{ij}$, $\min(\lambda_2 + \lambda_3 + \lambda_4)$.
Thirdly, we propose to use the locally
optimal probability matrix Eq.(~\ref{eq:stoch}) as transition matrix (solution
C) for the scattering of the worm at a vertex.\\ 

\begin{figure}
\includegraphics[width=8.5cm]{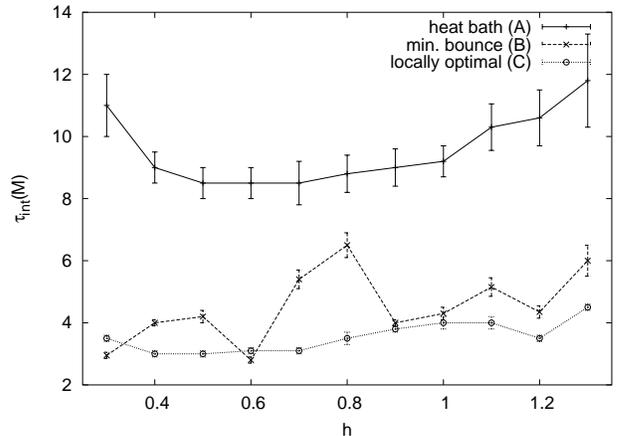}
\caption{\label{fig:spin_mag} Integrated autocorrelation times for the
  magnetization, $\tau_{int}(M)$, as a function of magnetic field $h$ for a spin $3/2 XY$ model as in
  Ref.~\cite{Alet03}, lattice size $L = 64$, inverse
  temperature $\beta = 64$. The integrated autocorrelation
  times are made loop size independent (normalized to two worms per update) so
  that the heat bath (solution A), minimal bounce (solution B) and locally
  optimal (solution C) algorithms can directly be compared. The precise
  definition of these algorithms is explained in the text.}  
\end{figure}

\begin{figure}
\includegraphics[width=8.5cm]{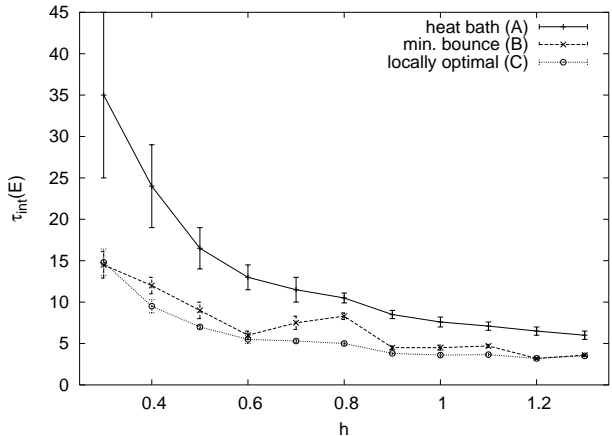}
\caption{\label{fig:spin_energy} Analogous to Fig.~\ref{fig:spin_mag} but
  now for the integrated autocorrelation time of the energy, $\tau_{int}(E)$,
  as a function of the magnetic field $h$.}
\end{figure}

{\it Spin $3/2 XY$ model.}
In Ref.~\cite{Alet03} the directed loop algorithm was studied for
the one dimensional spin $3/2 XY$ model in an external magnetic
field $h$,
\begin{equation}
H = J\sum_{\langle i,j \rangle} \frac{1}{2}({S}^{+}_i{S}_j^{-} +
{S}^{-}_i{S}_j^{+}) -h \sum_i S_i^z,
\end{equation}
where the first sum extends over nearest neighbors and $J$ is an exchange
interaction term. It appeared that solution B always gave shorter
autocorrelation times than solution A, as can also be seen in
Fig.~\ref{fig:spin_mag} and in Fig.~\ref{fig:spin_energy}, where the
integrated autocorrelation times for the uniform magnetization and the energy
are plotted. 
 
The authors of Ref.~\cite{Alet03} also proposed to break detailed balance to
$f_i W_i T_{ij} = f_j W_j T_{ji}$, with $f_i$ an extra degree of freedom at
leg $i$. Using a linear programming~\cite{NumRec} technique they were able to
further minimize the trace of the probability matrix. They found that this
algorithm gave the shortest autocorrelation times for most values of the
magnetic field, except around $h \approx 0.8$ where it behaved worse than
solution B and, unexpectedly, even worse than solution A. They deduced that
alternative principles than minimizing bounces might exist. Furthermore, this
algorithm needs to be used carefully, since it modifies condition (ii) with
respect to the invariant distribution for the Green's
function~\cite{Dorneich01} $\langle a_i^{\dagger}(0)a_j(t) \rangle$, although it is
correctly weighted with respect to the invariant distribution for diagonal
observables such as the energy and the magnetization. Note that it is also
possible to apply Eq.(~\ref{eq:stoch}) to the modified weights $W^\prime_i =
f_i W_i$ in order to obtain the locally optimal algorithm in case one is not
interested in the Green's function. Therefore we will not further compare this
proposal with solutions A, B and C.

We also addressed the spin $3/2 XY$ model with solution C and the results for
the integrated autocorrelation times for the magnetization and the energy are
also presented in Fig.~\ref{fig:spin_mag} and in
Fig.~\ref{fig:spin_energy}. We find a substantial improvement over solution
A. Solution C is also much better than solution B for magnetic fields around $h
\approx 0.8$, while for other magnetic fields the difference is smaller. This
indeed shows that minimizing bounces is not optimal for the spin $3/2 XY$
model. Note that increasing the system size does not significantly change the
ratio of the correlation times of solutions B and C. \\  
    
\begin{figure}
\includegraphics[width=8.5cm]{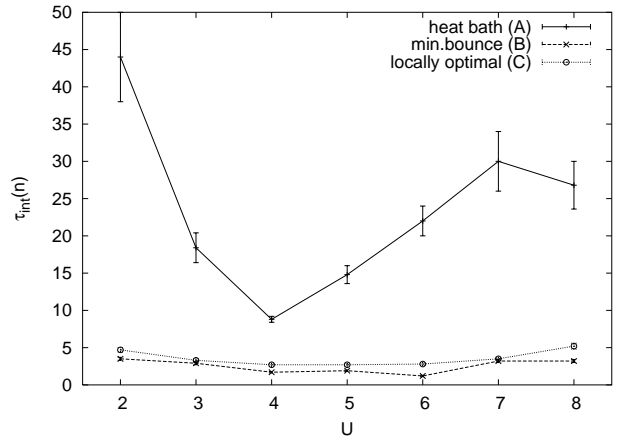}
\caption{\label{fig:bose_n} Integrated autocorrelation time for the density, $\tau_{int}(n)$, as a
  function of on-site repulsion $U$ for a one dimensional Bose-Hubbard model with
  $\mu = 5, t = 1$, lattice size $L = 64$ and inverse temperature $\beta =
  L$. Four loops were constructed in every update for the 'min. bounce (B)' and
  'locally optimal (C)' algorithms, while for 'heat bath (A)' we constructed 16 loops and
  multiplied the results by four afterwards. Particle number cut-off is
  lowered at $U = 3$ and $U = 8$. The Mott phase sets in for $U>9$.}
\end{figure}

{\it Bose-Hubbard model.} We also present results for the
Bose-Hubbard model~\cite{Zyubin04} in one dimension (units are as in Ref.~\cite{Pollet04}),
\begin{equation}
H = -t \sum_{\langle i,j \rangle} {b}^{\dagger}_i{b}_j +
 \frac{1}{2}U\sum_{i}{n}_i({n}_i-1)
 -\sum_{i} \mu {n}_i.
\end{equation}
The first term represents hopping with strength $t$ of the bosons between nearest neighbors, the
chemical potential is denoted by $\mu$ and the second term takes on-site
repulsion with strength $U$ into account. In the Bose-Hubbard model, the
diagonal weights are relatively much larger compared to the non-diagonal
weights than in spin systems. Again, as can be seen in Fig.~\ref{fig:bose_n},
the heat-bath updates (solution A) are outperformed, but for large on-site repulsion $U$
the minimal bounce solution (solution B) is superior to the locally optimal
solution (solution C). 

As in the Potts model, we are guaranteed that solution C is more efficient
than solution A, but the Peskun theorem does not claim that solution C is
superior to solution B, because choosing the lowest (local) $\lambda_2$ will not
necessarily correspond to the lowest integrated autocorrelation time,
$\tau_{int}$ (Eq.(\ref{eq:tauint}). Specifically, in case $n=2$ both solutions reduce to Metropolis
updating Eq.(\ref{eq:met}). In case $n=3$ solution B is of the form
\begin{equation}
T_{ij}^B = \left[
\begin{array}{ccc}
0 & a & 1-a \\
\frac{\pi_1}{\pi_2}a & 0 & 1 - \frac{\pi_1}{\pi_2}a \\
\frac{\pi_1}{\pi_3}(1-a) & \frac{\pi_2}{\pi_3} - \frac{\pi_1}{\pi_3}a &
\frac{2\pi_3 - 1 + 2 \pi_1 a}{\pi_3}
\end{array}
\right].
\end{equation}
The linear programming technique~\cite{NumRec} applied in solution B will try to make $a =
\frac{1-2\pi_3}{2\pi_1}$ if $\pi_3 < \frac{1}{2}$, otherwise it will take $a =
0$. It will be the system parameters that determine whether $T^B$ or $T^{C}$
(Eq.(\ref{eq:stoch}) for $n=3$) performs better. Due to its structure it is
also not possible to improve $T_B$ by Metropolizing it.

Also in case $n=4$ both solutions B and C put all diagonal elements in the
(local) stochastic matrix zero, except for the diagonal element corresponding
to the largest weight. We know from the Peskun theorem that
this leads to an efficient sampling. 
As a counterexample, the Metropolized Gibbs sampler Eq.(~\ref{eq:Liu}) has
only one zero on the diagonal of the transition matrix and is systematically
outperformed by both solution B and C.
 
The class of locally optimal stochastic matrices can be parameterized as
\begin{equation}
T_{ij} = \left[
\begin{array}{cccc}
0 & a & b & x \\
\frac{\pi_1}{\pi_2}a & 0 & c & y \\
\frac{\pi_1}{\pi_3}b & \frac{\pi_2}{\pi_3}c & 0 & z \\
\frac{\pi_1}{\pi_4}x & \frac{\pi_2}{\pi_4}y & \frac{\pi_2}{\pi_4}z &
1-\frac{\pi_1x + \pi_2y + \pi_3z}{\pi_4}
\end{array}
\right],
\end{equation}
with $ x = 1-a-b, y = 1-\frac{\pi_1}{\pi_2}a - c, z = 1-\frac{\pi_1}{\pi_3}b -
\frac{\pi_2}{\pi_2}c$ and the three free parameters $a, b$ and $c$. 
Solution B will now try to minimize $T_{44}$, or minimize $a, b$ and $c$,
under a number of constraints such as $a + b \leq 1$. This
can be suboptimal however, since the minimum will be found when one or
more of the constraints are exactly met~\cite{NumRec}. Suppose the minimum is found for $a+b
=1$, as it happens in the spin $3/2 XY$ model for magnetic fields $h
\approx 0.8$. Then the scattering from the least probable state to the most
probable state is zero, and this clearly cannot be optimal and explains why
solution C is better in Fig.~\ref{fig:spin_mag}.

The overall conclusion is that the integrated autocorrelation times for both
solutions B and C will be of the same order and roughly optimal. The important
principle is that the diagonal elements corresponding to the lowest $n-1$
weights should be zero. Some arbitrariness is still retained, but it does not
seem possible to define how the remaining freedom should be chosen
independently of the weights that occur in the updating process. When the
diagonal and non-diagonal weights are of the same order, solution C is better,
while for large diagonal weights solution B gives the lowest integrated
autocorrelation times. Furthermore, we have a good argument why these
solutions lead to a locally optimal sampling, based on the Peskun theorem. \\

{\it Conclusion.} We have shown that the optimal transition kernel for Markov
chain Monte Carlo should have a zero diagonal, except for the diagonal
element corresponding to the largest weight (which can be large).
We have presented results for the Potts model with random single spin updates
and for quantum spin chains and the Bose-Hubbard model with the directed loop
algorithms. They confirm the theoretical reasoning. Our results suggest a
practical way to improve existing Monte Carlo methods. This could lead to
significant gains in efficiency in both classical and quantum Monte
Carlo. One could consider applications in the research fields of flat histogram
methods~\cite{Wang01}, the loop algorithm~\cite{Evertz93}~\cite{Evertz03}, the
worm algorithm~\cite{Prokofev96} and the fast updates in auxiliary field
quantum Monte Carlo~\cite{Assaad01}~\cite{Rombouts99} and shell model Monte
Carlo~\cite{Koonin97}. Furthermore, our results could shed a
new light on the correlations in Glauber dynamics~\cite{Glauber63}.

The authors wish to thank the Research Board of the Universiteit Gent, the Fund
for Scientific Research, Flanders and NATO Grant PST-CLG 980420 for financial
support. The authors acknowledge G. Barkema, H. Bl\"ote, A. van Heukelum,
P. J. H. Denteneer, J. Carlson, K. Langanke, and J. Ryckebusch for valuable
discussions.   

\end{document}